\documentclass[11pt]{article}
\usepackage{moriond,epsfig}

\def\be{\begin{equation}}
\def\ee{\end{equation}}
\def\bea{\begin{eqnarray}}
\def\eea{\end{eqnarray}}

\def\beq{\begin{equation}}
\def\eeq{\end{equation}}
\def\bea{\begin{eqnarray}}
\def\eea{\end{eqnarray}}
\def\bq{\begin{quote}}
\def\eq{\end{quote}}

\def\tli{\Lambda^2}

\def\simlt{\stackrel{<}{{}_\sim}}

\begin{document}
\vspace*{4cm}
\title{ON GAUGE MEDIATION AND COSMOLOGICAL VACUUM SELECTION}

\author{ Zygmunt Lalak }

\address{Institute of Theoretical Physics, University of Warsaw\\
ul. Ho\.za 69, Warsaw, Poland}

\maketitle\abstracts{Gauge mediation of supersymmetry breakdown has many attractive features and can be realized in phenomenologically interesting string-motivated models. We point out that in models with the Polonyi-like field stabilized at a low expectation value by quantum corrections, gravity seems to limit from above the admixture of gravity mediation to the dominant gauge mediation channel. However, we also point out that in a class of  such models the low energy metastable supersymmetry breaking vaccum appears to be cosmologically disfavoured. These features should hold also in the case of typical stabilized models with anomalous $U(1)_A$ groups. }

\section{Introduction}
Gauge mediation of supersymmetry breaking 
\cite{GMfirst0,GMfirst1} 
with  the gravitino mass in the GeV mass range appears 
to be  a phenomenologically interesting
and theoretically well supported possibility. 
Heavy gravitino as LSP is an interesting dark matter candidate, allowing for a high reheating
temperature needed for leptogenesis.   
However, to reliably study  gauge mediation one needs to take into account the complete theory, with the dynamical supersymmetry breaking sector coupled to messengers and further to the visible sector. 
A simple and calculable supersymmetry breaking sector is given by the  
O'Raifeartaigh-type models, see, e.g.~\cite{OR,Kitano,rozne1,EOGM,Heckman:2008rb}.
The renewed interest in these models  is partly motivated by the acceptance of metastable supersymmetry
breaking vacua in models with $R$-symmetry broken spontaneously and/or
explicitly by a small parameter. 
There are two interesting aspects of such scenarios related to gravity. 
First of all, the gravitational mediation is always present and it is legitimate to ask to what extent one can mix the two channels of mediation. 
Secondly, one should ask about the cosmological history of the such models. The existence of many competing vacua - spersymmetric and non-supersymmetric ones - poses the question of how natural it is for the complete theory to settle down into the phenomenologically relevant 
vacuum with broken supersymmetry. We are going to make a few comments on these issues. This summary is based on~\cite{LPT,LD}.

\section{Models of direct gauge mediation}\label{subsec:prod}
A simple example is  a model 
discussed in \cite{Kitano} with superpotential containing a linear term of 
a gauge singlet chiral superfield $X$ responsible for supersymmetry  
breakdown.  The messengers $Q$ and $q$, transmitting the supersymmetry 
breakdown to the visible sector,     
transform as $\mathbf{5}$ and $\mathbf{\bar 5}$ of $SU(5)$ 
and are coupled to the field $X$. The superpotential reads   
$W = FX-\tilde\lambda XQq$ and the form of the K\"ahler potential,
\begin{equation}
\label{kitanokaehler}
K = \bar{X}X-\frac{(\bar{X}X)^2}{\tli}+\bar{q}q+\bar{Q}Q\, ,
\end{equation}
takes into account the loop corrections representing the logarithmic 
divergence in the effective potential coming from  the effects of the 
massive fields in the O'Raifeartaigh model which have been integrated out. 
The sign of the second term in the K\"ahler potential is negative here.
For $\tilde\lambda=0$ supersymmetry is broken by $F_X\neq0$ and $X$ is 
stabilized at $0$.  Supersymmetry is however restored by turning on 
the coupling $\tilde\lambda$. The supersymmetric global minimum is at 
$X=0$ and $Q=q=\sqrt{F/\tilde\lambda}$. Coupling the model to gravity changes 
the vacuum structure. Supersymmetric vacuum is still present as a global 
minimum (with shifted values of the fields) but in addition a local 
(metastable) minimum with broken supersymmetry   
appears, with  vanishing vevs for the messenger fields.

A constant $c$ is added to the superpotential to cancel the 
cosmological constant at the metastable vacuum. 
Another fact worth mentioning is that, with gravity, after decoupling 
the messengers ($\tilde\lambda=0$) the supersymmetric vacuum disappears 
(as in the case without gravity) but the minimum is for $X$ different 
from zero. 
In the above discussion, the role of gravity is linked to the 
negative sign of the second term in the K\"ahler potential, which 
follows from the O'Raifeartaigh model. However, more general models 
of a similar type can give positive sign for that term and $X$ can be 
stabilized away from the origin, with broken supersymmetry \cite{Shih},  
with or without messengers  even in the limit $M_P\to\infty$.  
One can expect that the role of gravity is then  more subtle.

\section{Graviational vs gauge mediation}
We shall now discuss the vacuum structure of a class of globally 
supersymmetric models under the additional assumption that  messengers  becoming massless in the limit $X\rightarrow 0$ give negligible 
contribution to the loop-corrected K\"ahler potential.
We expand the loop correction to the 
K\"ahler potential in powers of $|X|$:
\begin{equation}
\delta K = -\frac{m^2}{128\pi^2} \left ( f_4 (\frac{2 \lambda |X|}{ m})^{4} + (\frac{2 \lambda |X|}{ m})^{6}
 + \cdot \cdot \cdot \right ) \, .
\label{eff2a}
\end{equation}
In (\ref{eff2a}) we have extracted the overall dependence on the
representative mass scale $m$ and coupling strength 
$\lambda$.
 The scale $\Lambda$ appearing in
eq.\ (\ref{kitanokaehler}) is now given (for $f_4>0$) by
$\Lambda^2  = 8 \pi^2 m^2 /(\lambda^4 f_4)$ 
and it can be significantly larger than the scale $ m$, which is the mass scale of the rafertons giving rise to the domninant look correction.
In this expansion we denoted by $\ldots$ terms of higher order in $|X|$
as well as contributions from the messenger sector. We also 
suppressed the effects of the $|X|^0$ and $|X|^2$.
The former amounts to overall rescaling of the potential and the 
latter is swallowed by the rescaling of $X$ which restores its canonical normalization. 

Let us assume the simple 
superpotential:
\begin{equation}
W = m\phi_1\phi_3+\frac{R}{2}m\phi_2^2+\lambda \phi_1\phi_2 X +FX+c\, .
\label{wshih}
\end{equation}
A model described by (\ref{wshih})  has been extensively studied in 
the global limit in \cite{Shih}. Here we couple it to gravity
 and we again assume that the messenger sector does not affect the 
position of the local supersymmetry breaking minimum.
When $c=FM_P/\sqrt{3}$ the effective potential
vanishes in the limit $|X|\to 0$.
Only small corrections to this relation are necessary to make the
cosmological constant vanish at the supersymmetry breaking local minimum 
of the potential with $X\neq0$ and we shall use this
approximate relation from now on.

Functions $f_4$ and $f_6$, defined in eq.\ (\ref{eff2a}), 
have the following form:
\begin{eqnarray}
f_4 &=& -\frac{1+2R^2-3R^4+R^2(R^2+3)\ln R^2}{(R^2-1)^3} \\
f_6 &=& \frac{1+27R^2-9R^4-19R^6+6R^2(R^4+5R^2+2)\ln R^2}{3(R^2-1)^5} \, .
\end{eqnarray}
As shown in \cite{Shih}, the function $f_4$ is positive for $R<2.11$ 
and negative otherwise. 
The function $f_6$ is positive for $R>1/2$, thereby ensuring
(in the global limit) the existence of a metastable supersymmetry 
breaking minimum whenever $f_4<0$.

One can envision three main classes of 
solutions depending on the sign and size of $f_4$.
For $f_4<0$ and $f_6>0$ we recover the minimum previously discussed
in the context of globally supersymmetric model:
\begin{equation} 
X^2 = \frac{8|f_4|}{9f_6}\frac{ m^2 /4 }{ \lambda^2} \, .
\label{ssol1}
\end{equation}
For $f_4>0$ the position of the supersymmetry breaking minimum is 
determined by a balance between terms linear and quadratic in $X$
\cite{Kitano}. The solution reads then:
\begin{equation}
X = \frac{1}{2\sqrt{3}}\frac{\Lambda^2}{M_P} \, ,
\label{ksol}
\end{equation}
where, as before, $\Lambda$ appearing in (\ref{kitanokaehler})
is given by$\Lambda = 2 \pi  m/(\lambda^2n_\phi^{1/2}|f_4|^{1/2})$.
Finally, we may have $f_4\approx 0$, leading to a dominance of the
quartic term over the quadratic one.
We find then:
\begin{equation}
X^3  =  \frac{16\pi^2}{9\sqrt{3}/2  f_6} \frac{m^4}{16 \lambda^6M_P}\, .
\label{ssol2}
\end{equation}
Note that including the supergravity corrections to the effective
potential is crucial for the existence of
solutions (\ref{ksol}) and (\ref{ssol2}),
for which $\langle X\rangle$ is proportional to negative powers of $M_P$. 
All the three solutions (\ref{ssol1})-(\ref{ssol2})
break $R$-symmetry \cite{Nelson}: 
in (\ref{ssol1}) $R$-symmetry is broken spontaneously, whereas the form of 
(\ref{ksol}) and (\ref{ssol2})
shows that explicit soft $R$-symmetry breaking (the constant term $c$ in 
the superpotential) is transmitted to $\langle X\rangle$ through 
gravitational interactions.
Numerical analysis shows that all the three solutions can be realized in the simple model (\ref{wshih}),
depending on the value of the mass ratio $R$.

We note that the local minimum disappears for mass scales
of the  O'Raifeartaigh sector slightly smaller than $10^{-3}M_P$
This can be understood by noticing that the solution (\ref{ssol2}),
which is a good approximation for sufficiently large $ m$,
can be rewritten as:
\begin{equation}
\left(\frac{\lambda X}{m/2}\right)^3 = \frac{32\pi^2}{9\sqrt{3}f_6}\frac{m}{2 \lambda^3 M_P^3}
\label{ssol22}
\end{equation}
If the left-hand side of (\ref{ssol2}) exceeds unity, our perturbative 
expansion (\ref{eff2a}) breaks down and one should not expect the minimum
to persist. It also follows from (\ref{ssol22}) that the maximal
scale ${m}$ for which there exists a local minimum with $X\neq0$
scales as $\lambda^3$, which, upon substitution to (\ref{ssol2}) 
shows that the corresponding value of $X$ scales as $\lambda^2$.
These observations are confirmed by our numerical analysis.

The numerical analysis supports the conclusion that  the 
supergravity corrections
provide an upper bound on the values of the
mass scale $m$ of the O'Raifeartaigh sector for which
the metastable supersymmetry breaking minima exist.
In the particular example analyzed here,this bound is 
$m\simlt 10^{-3}M_P$, which corresponds to 
$\Lambda\simlt 10^{-2}M_P$.

\section{Cosmological vacuum selection}

To describe the cosmological history of models with direct gauge mediation let us restrict ourselves to the case $f_4 =1, \; f_6 =0$. The result is rather generic as confirmed by more general analysis~{\cite{LD}}. Let us assume that the hidden sector coupled to messengers is in thermal equilibrium shortly after inflation. This is justified as the interactions of $X$ with messengers and with the observable sector are not suppressed by a large mass scale, like in the case of the purely gravitational mediation. It is rather straightforward to see that at very hight temperatures, $T \ll \tilde{\Lambda}$ the minimum of the effective potential lies very close to the origin in the field space. $Q=q=0, \; X \approx 0$. As the Universe cools down, at the critical temperature
\beq
T_{cr} = 2 \frac{\mu}{\sqrt{\lambda}} \; , 
\eeq 
the two minima with nonvanishing expectation values of messengers, $<q>=<Q>\neq 0$, form, which evolve smoothly towards the supersymmetric minima at $T=0$. The mimimum which corresponds to the supersymmetry breaking minimum at low temperatures forms at the temperature $T_X$ which is typically much lower than $T_{cr}$:
\beq
T_{X} = {\cal O}(3) \, \mu \frac{\mu}{\Lambda \lambda} \approx T_{cr} \frac{1}{\sqrt{\lambda}} \frac{\mu}{\Lambda}\; .
\eeq
The possible way out is to arrange for nonadiabatic initial conditions, which give rise to a dispaced intial value for the field $X$. 
Analysis of the dynamical evolution indicates, that a suitable set of iniatial conditions leading to enhanced probability of the evolution towards the non-supersymmetric vacuum is $\Lambda^2 < X_{init} < \Lambda$ with $ \lambda < 10^{-7}, \; 10^{-3} < \Lambda < 10^{-1}$. Interestingly, this points towards the mixed gauge/gravity mediation scenario. 

\section{Summary}\label{subsec:fig}
Gauge mediation of supersymmetry breakdown has many attractive features and can be realized in phenomenologically interesting string-motivated models. We point out that in models with the Polonyi-like field stabilized at a low expectation value by quantum corrections, gravity seems to limit from above the admixture of gravity mediation to the dominant gauge mediation channel. However, we also point out that in a class of such models the low energy metastable supersymmetry breaking vaccum appears to be cosmologically disfavoured. Our conclusions are not altered in the models where $X$ is charged under  an anomalous $U(1)_A$ group, and coupled to charged modulus, as long as the modulus and the gauge boson of  the $U(1)_A$ obtain the Planck scale masses, which is usually the case.

\section*{Acknowledgments}
\noindent This work was partially supported by the 
EC 6th Framework Programme MRTN-CT-2006-035863,  
and  by TOK Project  
MTKD-CT-2005-029466. 
ZL thanks CERN Theory Division for hospitality.

\end{document}